\documentclass[preprint,12pt,sort&compress]{elsarticle}

\usepackage{amssymb}
\usepackage{amsmath}

\usepackage{booktabs}
\usepackage{multirow}
\usepackage{subcaption}
\usepackage[unicode=true,
bookmarks=true,bookmarksnumbered=true,bookmarksopen=true,bookmarksopenlevel=2,
breaklinks=false,pdfborder={0 0 1},backref=false,colorlinks=true]
{hyperref}

\usepackage{algorithm}
\usepackage{algpseudocodex}

\usepackage{bbm}

\usepackage{array}
\newcolumntype{P}[1]{>{\centering\arraybackslash}p{#1}}

\newcommand{\bfA}{\mathbf{A}}
\newcommand{\bfB}{\mathbf{B}}

\newcommand{\bfR}{\mathbf{R}}

\newcommand{\bfZ}{\mathbf{Z}}
\newcommand{\bfi}{\mathbf{i}}
\newcommand{\bfj}{\mathbf{j}}
\newcommand{\bfl}{\mathbf{l}}
\newcommand{\bft}{\mathbf{t}}
\newcommand{\bfn}{\mathbf{n}}

\newcommand{\ep}{\varepsilon}

\journal{Mathematics and Computers in Simulation}

\begin{document}
	
\begin{frontmatter}

\title{Apictorial Jigsaw Puzzle Reconstruction Based on Curve Matching via a Corotational Beam Spline}

\author[1]{Igor Orynyak}
\ead{igor_orinyak@yahoo.com}
\author[1]{Dmytro Koltsov}
\ead{koltsovdd@gmail.com}
\author[1]{Danylo Tavrov\corref{cor1}}
\ead{tavrov.danylo@lll.kpi.ua}
\cortext[cor1]{Corresponding author}

\affiliation[1]{organization={Applied Mathematics Department, Igor Sikorsky Kyiv Polytechnic Institute}, 
	addressline={37 Beresteiskyi Ave.}, 
	city={Kyiv},
	postcode={03056},
	country={Ukraine}}

\begin{abstract}
Automatic assembly of apictorial jigsaw puzzles presents a classic curve matching problem, fundamentally challenged by discrete and noisy contour data obtained from digitization. Conventional smoothing methods, which are required to process these data, often distort the curvature-based criteria used for matching and cause a loss of critical information. This paper proposes a method to overcome these issues, demonstrated on the automatic reconstruction of a 54-piece puzzle.

We reconstruct each piece's contour using a novel corotational beam spline, which models the boundary as a flexible beam with compliant spring supports at the measured data points. A distinctive feature is the dynamic re-indexing of these points; as their calculated positions are refined, they are re-numbered based on their projection onto the computed contour.

Another contribution is a method for determining spring compliance in proportion to the distance between the point projections. This approach uniquely ensures a uniform degree of smoothing for corresponding curves, making the matching process robust to variations in point density and dependent only on measurement accuracy. Practical computations and the successful automatic reconstruction of the puzzle demonstrate the proposed method's effectiveness.
\end{abstract}

\begin{highlights}
	\item A corotational beam spline reconstructs contours from noisy, discrete data
	\item Adaptive spring rigidity ensures uniform smoothing independent of point density
	\item Data points are renumbered based on their projection onto the computed contour
	\item The method processes data where measurement error exceeds point distance
	\item Optimal smoothing maximizes the energy difference between correct and incorrect matches
\end{highlights}


\end{frontmatter}

\section{Introduction}

Determining the correspondence between segments of two curves measured with some degree of error is a fundamental problem in pattern recognition. This task plays a crucial role across many areas of computer engineering. Examples include the reconstruction of archaeological artifacts from fragments---such as mosaics, pottery, jewelry, and skeletal remains \cite{mcbride2003,kleber2009}; the restoration of intentionally or accidentally damaged documents and banknotes \cite{justino2006,yilmaz2023}; the detection of artificial insertions in digital images \cite{cho2010}; the analysis of DNA structures in biology \cite{marande2007,markaki2023}; the recognition of natural objects \cite{gope2005}; and the reconstruction of the causes and patterns of structural or mechanical failures \cite{leitao2005}. All of these studies are, in one way or another, related to the problem of automatically assembling apictorial two-dimensional jigsaw puzzles. As Freeman noted, this “represents a well defined and easily understood problem,
and since it aptly illustrates the difficulties besetting all pattern recognition problems,
it is ideally suited as a guinea pig for testing new pattern recognition techniques” \cite{freeman1964}. This observation aptly characterizes the motivation behind the present study.

This problem is non-trivial due to both computational complexity and methodological challenges in geometric curve matching \cite{goldberg2002}. The complexity arises from the large number of elements and possible matching segments. Key methodological difficulties include selecting a suitable matching criterion—such as point distances, tangent angles, or curvatures—and its threshold, as well as addressing measurement inaccuracies caused by instruments, object properties, and environmental factors.

Measurement inaccuracies make direct curve matching impossible, a problem exacerbated by matching criteria like angles and curvatures, which, as derivatives, amplify high-frequency noise during numerical differentiation. While smoothing can suppress this noise, it risks removing parts of the useful signal. The challenge is heightened when matching two curves measured under different conditions (e.g., varying point density or noise levels). 

In the following review of the literature, we will focus primarily on issues related to curve smoothing and the use of curvature as a criterion for line similarity, while paying less attention to the specific sequence of element assembly, which has been examined in considerable detail in previous studies.

The first study addressing the ``assembly'' of apictorial puzzles was that of \cite{freeman1964}. In addition to formulating the problem itself, the author identified three main aspects: (1) digital representation of element boundaries; (2) manipulation of elements, including rotation and matching; and (3) evaluation of boundary correspondence. The matching process relied on identifying points where curvature was extreme. The issue of smoothing, however, was not considered.
At that time, it was common to perform curve matching based solely on characteristic points, while smoothing was mostly limited to specific methods for interpreting whether a given pixel belonged to an element's region, depending on the state of neighboring pixels \cite{levine1969}. As for contour smoothing, subjective criteria were used to assign weights to successive points relative to preceding ones \cite{levine1969}.

A significant contribution was made by Wolfson et al. \cite{wolfson1988}, who examined a rectangular (four-sided) jigsaw puzzle consisting of 104 pieces. The study described a method for digitally photographing the puzzle and identifying the four corner points, corresponding to points of high curvature.
For the first time, the need for contour smoothing was explicitly recognized, following the idea of \cite{schwartz1987}, where the shortest path between measured points is found, constrained to lie within the contour's known error margin, and then re-sampled with equidistant points.
To determine whether two curves correspond, the Euclidean distance between their respective points is calculated \cite{wolfson1988}, and the correspondence is assumed to be optimal when this distance is minimal.

The subsequent work by Wolfson \cite{wolfson1990} was the first to introduce the dependence of curvature on the arc-length coordinate as the most informative characteristic of a curve. At the time, curvature was approximated using locally averaged values of angular increments over elementary segments of a given length. In \cite{kurnianggoro2018}, a review is provided of state-of-the-art methods for representing the shapes of two-dimensional figures and the key parameters (features) that characterize them, with curvature again being one of the most significant (distinguishable) features. 

A similar technique is applied in the reconstruction of cut or torn documents \cite{stieber2010}, where the correspondence between two contours is verified for the entire set of points by computing the sum of coefficients that are predefined according to the differences of squared curvatures.
A more general approach to comparing two contours obtained at different scales was proposed in \cite{duncan1991} and \cite{cohen1992}, where the notions of deformation energy and stretching energy were introduced---defined as the integrals of the squared differences of curvatures and the squared differences of linear deformations (stretching), respectively. These comparison criteria were later used in studies of archaeological, biomedical, and graphical images \cite{mcbride2003,sebastian2000,sebastian2003}.

It is interesting to note that similar integrals for a single curve are widely used in interpolation problems and in constructing optimal paths or aesthetic curves. They have a clear mechanical interpretation, since the deformation curve of a flexible beam satisfies the criterion of minimum deformation energy \cite{levien2009,brunnett1994,orynyak2025b}. A similar criterion will be applied in this paper.

Methods for smoothing and obtaining continuous curvature values for discretely defined position points were proposed earlier, in 1986 \cite{mokhtarian1986}. It is assumed that the smoothed coordinates of the points $x$ and $y$ can be represented as functions of an independent parameter $t$. The values $t_i$ at the measurement points are assumed known---for example, they may be equal to number $i$---and the measured positions are denoted by $X_i$ and $Y_i$, respectively. To obtain continuous smoothed dependencies $x(t)$ and $y(t)$, a convolution integral is introduced:
\begin{equation}
	(x(t); y(t)) = \int_{-\infty}^{+\infty} \left(X_i(u); Y_i(u)\right) g(t-u,\ \sigma)\,du\;,
    \label{eq:ConvolutionIntegral}
\end{equation}
where kernel $g(t, \sigma)$ is the normal (Gaussian) density with variance $\sigma^2$. Clearly, $\sigma$ serves as the smoothing parameter (bandwidth), and the larger it is, the stronger is the smoothing. Having obtained continuous values of the functions $x(t)$ and $y(t)$, one can determine the arc-length coordinate $s(t)$ and the curvature $\kappa(t)$ of each point.

It should be noted that Gaussian smoothing is the most commonly applied method in the processing of measurement data \cite{goldberg2004,michel2011}.
However, in \cite{mokhtarian1986}, the choice of the parameter $\sigma$ in relation to the amplitude of possible measurement errors is not specified; only general remarks are given regarding its influence on the resulting curve.
In statistical estimation, a similar approach is known as the Nadaraya-Watson method, which is not considered optimal for smoothing \cite{eubank1999}.
In particular, to account for unevenly spaced points $X_i$ and $Y_i$, and when the distance between them is comparable to the smoothing parameter $\sigma$, certain interpolation methods are proposed to estimate intermediate values for use in the convolution integral (\ref{eq:ConvolutionIntegral}) \cite{clark1977,cheng1981}.

Moreover, the use of parametric splines or kernels in geometric problems depends on the subjective choice of the parameter $t$, such as a uniform distribution or one linked to the arc length (the so-called natural parametrization) of the interpolation spline \cite{moreton1992}.
However, neither of these popular approaches is optimal, and alternative methods are often employed---in particular, the so-called centripetal method, where the value of $t$ depends on the curvature of the trajectory itself \cite{lee1989}.
Like natural parametrization, it is iterative, but it produces better results in regions with sharp curvature changes.

An alternative smoothing approach is used for the identification of marine mammals \cite{gope2005}.
Instead of the above-mentioned convolution integral---which requires a large number of computations and poses difficulties with boundary conditions for finite segments---the B-spline technique is employed \cite{gope2005}.
Here too, a parametric approximation is used, and a coordinate, for example $x(t)$, is found by minimizing a functional that, in addition to the least squares deviations, includes a term that minimizes the integral of the squared curvature with a certain weight $\lambda$ (minimization of a penalized least squares problem):
\begin{equation}
	E(\lambda) = \sum_{i}\left(x(t_i) - X_i\right)^2 + \lambda\int_{-\infty}^{+\infty} \left(\ddot{x}(t)\right)^2\,dt\;.
    \label{eq:MinimizationIntegral}
\end{equation}

The smoothing parameter $\lambda$ plays a role similar to that of the parameter $\sigma$. Analogous approximations are widely used in statistical analysis \cite{ruppert2002}. It is worth noting that cubic (beam) splines minimize the bending energy for each chosen value of the proportionality coefficient between the increment of the third derivative and the function at the given discrete points \cite{reinsch1967}.

The problem of noise is identified as a central issue in pattern recognition in \cite{manay2006}. It is noted that the successful application of differential invariants is hindered by the complexity of their computation, which amplifies noise. For this reason, integral equivalents of these invariants are proposed. In \cite{cui2009}, smoothing is performed using B-splines, and an integral of the modulus of curvature is introduced as a characteristic for comparing curves or their segments. All of this underscores the lack of reliable and interpretable methods for smoothing curvature to a clearly defined level.

The original affine-invariant curve descriptor for curve matching and hidden object recognition \cite{fu2013} also requires the determination of curvature, followed by integration of its absolute values. The smoothing procedure occurs in two stages. In the first stage, a Gaussian kernel is applied, although, as noted in \cite{fu2013}, the choice of the parameter $\sigma$ is ambiguous: small values yield many points where the curvature changes sign, while large values lead to a loss of useful signal. In the second stage, B-splines are used to generate a continuous and further smoothed contour suitable for obtaining continuous curvature values.

The authors of \cite{hoff2013,hoff2014} use curvature and its derivative with respect to the contour length as Euclidean differential invariants. Characteristic points are defined as those where the derivative of the curvature reaches zero. These points are used to segment the curves into regions for curve matching.
A significant influence of noise on the obtained curvature values is noted, which requires rather precise measurements of the contour \cite{hoff2013,hoff2014}. While this reduces the noise problem, it does not eliminate it entirely. Two empirical smoothing methods are proposed. The first method involves shifting points of maximum curvature in the direction of the curvature by an amount proportional to the curvature itself. This is a subjective method, which may smooth two corresponding curves differently, complicating the matching procedure.
The second method is also subjective and uses a periodic spline technique with an iterative procedure that redistributes the initial points proportionally along the spline length, with the number of iterations reaching up to 1500 \cite{hoff2014}. As noted by the authors, this method lacks a theoretical justification, and they propose using smoothing splines in future work.

A key aspect in all methods for reproducing, defining, and performing shape matching is the smoothing of the primary measurement data. Similar contour smoothing problems arise in related fields of computer graphics, particularly for constructing trajectories or paths from imprecisely measured points.
Often in such problems, the distance between measurement points significantly exceeds the measurement error; for example, in \cite{gwon2017}, the distance is 1 meter, while the error is 0.05 m. This allows for straightforward numbering of points (even if they are not time-stamped) and processing them using, for instance, B-splines.
The work \cite{bertolazzi2020} is noteworthy because, although the authors use ordered data, the observed scatter is large enough that, during the spline construction process, the points must be renumbered. In extreme cases, their projections onto the spline appear in a different order than the original indices. The spline construction algorithm consists of several steps, including clustering of points, construction of cubic polynomials for each cluster, and then applying the classical cubic parametric spline to averaged cluster polynomials.
%
%
%
Evidently, this is a complex methodology with subjective steps and no mechanism for controlling curvature, which in the examples exhibits significant fluctuations \cite{bertolazzi2020}.

Our work is devoted to the application of corotational beam splines (CBS) for controlled smoothing---that is, smoothing in which noise is reduced to a clearly defined, acceptable level for subsequent processing---of curvature in geometric figures from imprecisely measured data. 

This work's primary aim is to develop a reliable method for extracting curvature from noisy contour data via CBS. Its novelty lies in addressing scenarios with densely sampled points where measurement error---potentially several pixels due to lighting or color ambiguities---exceeds the distance between the points themselves. This high-noise condition necessitates iterative procedures and point renumbering. Therefore, the smoothing process must be sophisticated enough to preserve a high signal-to-noise ratio while applying an equivalent degree of smoothing to both curves, regardless of initial measurement differences. Consequently, the focus is not on puzzle-solving performance but on developing robust contour-matching criteria and methods for the controlled smoothing of artifacts. 

The theory and practice of applying CBS are described in our works \cite{orynyak2025,orynyak2024,orynyak2022,orynyak2025b}. CBS represents a development of the well-known beam interpolation splines, which trace back to \cite{holladay1957}. For smoothing tasks, cubic beam splines were first applied in \cite{mehlum1964} as a minimization of a functional of the form \eqref{eq:MinimizationIntegral}, where the coordinate $y$ is treated as a function of the coordinate $x$. Interestingly, this functional is essentially the solution to the differential equations of beam theory, where at the measurement points (supports), forces are in action, which are proportional to the difference between the computed value $y(x_i)$ and the measured value $y_i$, with a proportionality coefficient $\lambda$ \cite{mehlum1964}. Similar approaches are still widely used in statistical analysis, with their origins in Reinsch's work \cite{reinsch1967}.

In geometric analysis, solutions of the form $y(x)$ cannot generally be applied, since the two coordinates are essentially treated unequally. Thus, an explicit function $y(x)$ allows spline construction only when the tangent direction changes by no more than $180^\circ$. There are two ways to extend the use of geometric splines. The first is to treat both coordinates as functions of an independent parameter $t$, i.e., $y(t)$ and $x(t)$, as proposed in \cite{ferguson1964}, with subsequent optimization of the parameter $t$ \cite{kjellander1983,lee1989}.

Another way to apply beam spline ideas is to treat the curve equation as an implicit function, i.e., $g(x, y)=0$, as proposed by Fowler and Wilson in \cite{fowler1966}. Their main idea was to use local coordinates for each segment connecting two consecutive points. For spline construction, a smoothing condition was applied to the angles between consecutive segments. The main limitation of this approach is that the angular distance between segments must not exceed $8^\circ$--$10^\circ$ \cite{birchler2001}, and these splines were not applied to approximation tasks.


The rest of the paper is as follows. In Sect. 2, we briefly review the CBS construction method, with a special emphasis on adaptive spring support rigidities. Section 3 provides details on how this general approach is tailored to the specific problem of assembling apictorial jigsaw puzzles. In Sect. 4, we illustrate our technique using a real puzzle and discuss the challenges that arise in practice. Section 5 concludes.

\section{Corotational Beam Splines}

\subsection{Construction Method}

\begin{figure}
    \centering
    \includegraphics[width=\textwidth]{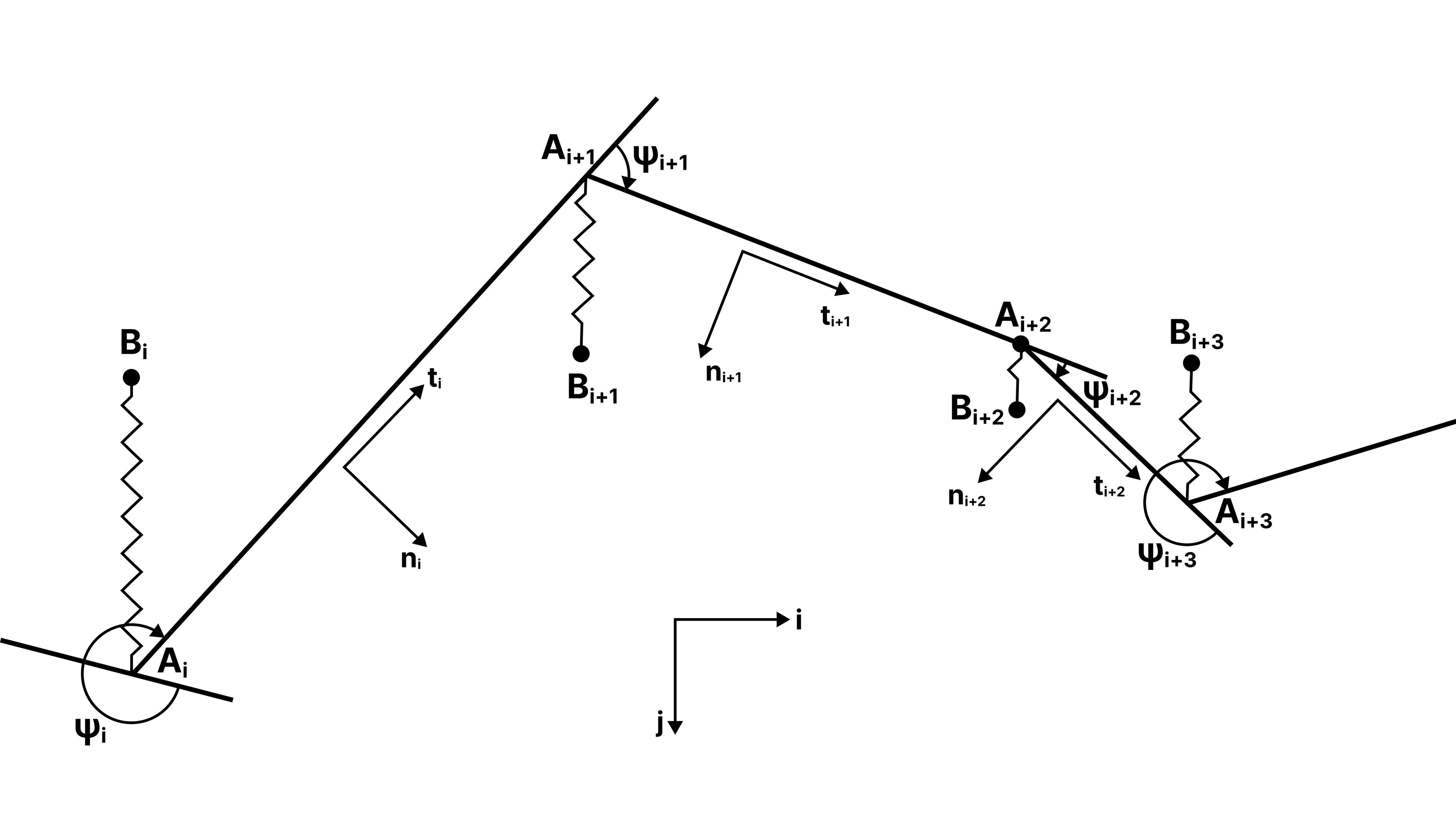}
    \caption{Model of discrete beams on elastic supports}
    \label{fig:Model}
\end{figure}

The CBS algorithm takes as input a set of ordered points (in our case, pixels) $\bfB_i(X_i,Y_i)$, $i = 1, \ldots, M + 1$, in a rectangular coordinate system, where $i$ is the point number in the dataset. At each iteration $k$, the points of the fixed initial set $\bfB_i$ are corresponded with calculated points from the approximated set $\bfA_i^k(X_i^k,Y_i^k)$ (Fig.~\ref{fig:Model}); hereafter, these points will be referred to as ``control points.'' The index $k$ will be omitted in most cases. All points, $\bfA_i$ from the calculated set, and $\bfB_i$ from the input set, are defined in the global coordinate system $(\bfi, \bfj)$.

We connect the points $\bfA_i^k$ and $\bfA_{i+1}^k$ sequentially with straight lines, forming segments. These segments will be denoted by the index $i$, corresponding to the starting point of the segment (Fig.~\ref{fig:Model}). The length of the vector $\bfl_i$ for each segment is determined by the following formula:

\begin{equation}
	\bfl_i = \bfA_{i+1} - \bfA_i = (X_{i+1} - X_i) \bfi + (Y_{i+1} - Y_i) \bfj\;.
    \label{eq:VectLen}
\end{equation}

For each segment, a local coordinate system $(s_i,w_i)$ and the corresponding basis vectors $\bft_i$ and $\bfn_i$ are defined. The local tangent vector $\bft$, based on formula (\ref{eq:VectLen}), is determined as follows:

\begin{equation}
	\bft_i = \frac{\bfl_i}{\left|\bfl_i\right|} = a_i \bfi + b_i \bfj\;,
	\label{eq:DecimalVector}
\end{equation}
where
\begin{equation}
	\left|\bfl_i\right| = l_i = \sqrt{\left(X_{i+1} - X_i\right)^2 + \left(Y_{i+1} - Y_i\right)^2}\;.
\end{equation}

The normal vector $\bfn_i$ is defined as perpendicular to $\bft_i$ and rotated clockwise relative to $\bft_i$ (Fig.~\ref{fig:Model}). The local coordinate system $(s_i,w_i)$ is related to the basis vectors as follows. The origin of $s_i$ coincides with $\bfA_i$ and its positive direction coincides with the vector $\bft_i$. Similarly for $w_i$, whose origin coincides with $\bfA_i$ and whose positive direction coincides with the vector $\bfn_i$. The vector $\bfn_i$ is determined as follows:
\begin{equation}
	\bfn_i = c_i\bfi + d_i\bfj\;,
    \label{eq:NormalVector}
\end{equation}
where
\begin{equation}
	\begin{pmatrix}c_i \\ d_i\end{pmatrix} = \begin{pmatrix}\cos\left(-\frac{\pi}{2}\right) & -\sin\left(-\frac{\pi}{2}\right) \\ \sin\left(-\frac{\pi}{2}\right) & \cos\left(-\frac{\pi}{2}\right)\end{pmatrix} \begin{pmatrix}a_i \\ b_i\end{pmatrix} = \begin{pmatrix}0 & 1 \\ -1 & 0\end{pmatrix} \begin{pmatrix}a_i \\ b_i\end{pmatrix}\;.
    \label{eq:RotateMatrix}
\end{equation}

Also playing an important role in the construction of the smoothed contour is the alignment of angles between two adjacent straight segments, i.e., between the directions $\bft_i$ and $\bft_{i+1}$. We denote these misalignment angles as $\psi_{i}$; their positive direction is clockwise (Fig.~\ref{fig:Model}), and they are calculated using inner product formulas:
\begin{gather}
    \cos(\psi_i) = \bft_i\cdot\bft_{i+1}\;, \\
    \sin(\psi_i) = \bfn_i\cdot\bft_{i+1}\;.
    \label{eq:Scalar}
\end{gather}

To determine the quadrant in which the angle lies, we need to determine both values, as individually these functions are only defined on two quadrants of the coordinate system. Note that if we get positive values in both equations, the angle is in the first quadrant. 

Let us now consider the basic mathematical model of the straight beam bending method, its parameters, and equations. The beam is described by a state vector $\bfZ(t)$, which is characterized by a set of four parameters for each local length coordinate $t$, measured from the starting point of the segment, $\bfZ(t) = (W(t); \theta(t); M(t); Q(t))^\top$. Here, following the concepts of beam theory, instead of the deflection function and its derivatives, we operate with the following quantities: $W(t)$ is the displacement, whose positive direction coincides with the direction of the normal; $\theta(t)$ is the angle of rotation, whose positive direction coincides with the direction of clockwise rotation from the tangent to the normal; $M(t)$ is the bending moment; and $Q(t)$ is the transverse force. Their positive directions are chosen so that a positive sign is used in the corresponding differential dependencies. Here we consider the simplest static beam, the so-called Euler-Bernoulli beam, which is described in all textbooks on the strength of materials. The following differential dependencies between the beam parameters are used:
\begin{gather}
    \frac{dW(t)}{dt} = \theta(t)\;,
    \frac{d\theta(t)}{dt} = M(t)\;,
    \frac{dM(t)}{dt} = Q(t)\;,
    \frac{dQ(t)}{dt} = 0\;.
    \label{eq:Dependences}
\end{gather}

The solution of the system of governing equations (\ref{eq:Dependences}) in a form convenient for applying the transfer matrix method is as follows:
\begin{equation}
	\bfZ(t) = \left[p_{i,j}(t)\right] \bfZ_0 \;,
    \label{eq:SimpleSolution}
\end{equation}
for $\bfZ_0 \equiv \bfZ(t=0)$, which is the state vector at the initial point of the considered segment, and
\begin{equation}
	\left[p_{i,j}\left(t\right)\right] = \left[
        \begin{matrix}1&t&\frac{t^2}{2}&\frac{t^3}{6}\\0&1&t&\frac{t^2}{2}\\0&0&1&t\\0&0&0&1\\\end{matrix}\right]\;, \quad 0\le t\le l_i\;.
    \label{eq:Matrix}
\end{equation}

To formulate the calculation scheme, we need to supplement the transfer matrix equation (\ref{eq:Matrix}) with conjugation equations, which relate the state vector at the end point of the previous segment to that at the starting point of the next one:
\begin{gather}
    W_{i+1,0} = W_{i,1}\;, \label{eq:Dependences1} \\
    \theta_{i+1,0} = \theta_{i,1} - \psi_{i}\;, \label{eq:Dependences2} \\
    M_{i+1,0} = M_{i,1}\;, \label{eq:Dependences3} \\
    Q_{i+1,0} = Q_{i,1} - D_i (W_{i,1} - \Pi_i)\;, \label{eq:Dependences4}
\end{gather}
where the subscripts 0 and 1 denote the beginning and end of the segment, respectively.
Note that if $D_i$ approaches infinity, which occurs with very small scatter in the input data (compared to the distances between measurement points) or when they are determined precisely (an interpolation problem), equation (\ref{eq:Dependences4}) cannot be implemented in a computational program. In this case, it should be represented as:
\begin{equation}
	W_{i,1} = \Pi_i - C_i(Q_{i+1,0} - Q_{i,1})\;,\quad C_i = \frac{1}{D_i}\;.
    \label{eq:Dependences4Alt}
\end{equation}

Equation (\ref{eq:Dependences1}) means that the displacement (deviation of the position from the initial straight line) must be the same. Equation (\ref{eq:Dependences2}) ensures the tangential continuity of the deformed contour, where the deformation angles compensate for the initial angular misalignment. Equation (\ref{eq:Dependences3}) is the equality of the approximate beam curvatures. And equation (\ref{eq:Dependences4}) takes into account the action of a spring support placed between the sections, where $D_i$ is the rigidity of the spring support (likewise, in \eqref{eq:Dependences4Alt}, $C_i$ is the compliance of the spring support). The concept of $\Pi_i$---the distance (gap) between the support (point $\bfB_i$) and the nearest point on the contour $\bfA_i$---deserves special attention. The search for and renumbering of points $\bfA_i$ at each iteration is a significant novelty of our paper; the corresponding procedure will be described below. 

Note that the system of equations (\ref{eq:Matrix}) and (\ref{eq:Dependences1}--\ref{eq:Dependences4}) (if the change in angle $\psi_{i}$ is not considered in the latter equations, which is natural for statistical splines) reduces to the following differential equation \cite{reinsch1967}:
\begin{equation}
	\frac{d^4W (t)}{dt^4} = D_i\delta(t - l_i) (\Pi_i - W(t))\;,
    \label{eq:Dirac}
\end{equation}
where $\delta$ is the Dirac delta function.

\subsection{Construction of the Resulting Spline}

The algorithm for constructing the initial polygonal contour at each iteration, numbering the unknowns, forming the system of equations, solving it to obtain the smoothed geometry, and transitioning to the next iteration which requires a polygonal geometry---all this is described in our previous works \cite{orynyak2025,orynyak2022,orynyak2025b}. Here, we will only note the following.

Before each iteration, we have numbered points $\bfB_i$ and $\bfA_i$, $i = 1, \ldots, M + 1$. The latter form a polygon with $M$ straight segments. From the positions of the  $\bfB_i$ and $\bfA_i$ points, we find the signed distance $\Pi_i$ between each pair of points, $i = 1, \ldots, M + 1$. The sign is positive if the vector $\overrightarrow{\bfA_i \bfB_i}$ forms an acute angle with the normal $\bfn_i$. 

For each segment, we form a vector of unknowns (4 parameters from the state vector $\bfZ_i(t)$) at both ends ($t = 0$ and $t = l_i$), totaling 8 unknowns. Since the number of segments is $M$, we have a total of $8M$ unknowns. On the other hand, for each segment, we have 4 transfer matrix equations \eqref{eq:SimpleSolution}, i.e., $4M$ equations. At the boundaries between elements (segments), we have 4 conjugation equations  \eqref{eq:Dependences1}--\eqref{eq:Dependences4}. There are $M-1$ such boundaries, giving us $4M-4$ equations. Also, 2 boundary conditions are applied at each end, for a total of 4. Thus, the number of equations and the number of unknowns are the same.

Solving this system of equations yields state vectors $\bfZ_i(0)$ and $\bfZ_i(l_i)$, $i = 1, \ldots, M + 1$. Using the state vectors at the beginning  of each segment, $\bfZ_{1,0}, \ldots, \bfZ_{M,0}$, we can apply \eqref{eq:SimpleSolution} to determine the state vectors $\bfZ_i(t)$ for each point $t$ of each segment, $i = 1, \ldots, M$.

To provide the continuity of spline near the ends of segment, in our CBS, we introduce the deformed normal vector,
 $\bfn^\theta_i(t)$, along which the calculated
 $W(t)$ is placed. This deformed
 normal vector $\bfn^\theta_i(t)$ is obtained
 by rotating the initial vector $\bfn_i$ by the calculated
 angle $\theta(t)$. It is easy to check that due to the
 continuity of angles \eqref{eq:Dependences2}, the deformed normal vectors would
 become continuous:
  \begin{equation}
	\bfn^\theta_i(t) = \bfn_i\cos\theta(t) - \bft_i \sin\theta(t)\;.
    \label{eq:NormalVectors}
\end{equation}

 The value of the position of the points of segment $i$ on the resulting spline
 is given as the vectorial sum of the initial position (straight segment)
 and the calculated displacement:
   \begin{equation}
	\bfR_i(t) = (\bfA_{i-1} + t\cdot \bft_i) + W_i(t)\bfn^\theta_i(t) \equiv \bfA_{i-1} + \bft_i\cdot T(t) + \bfn_i\cdot N(t)
    \label{eq:Displacement}
\end{equation}

Further analysis allows us to obtain the (deformed, spline) arc length coordinate, $s(t)$, and the curvature, $\kappa_d(t)$, at each point of the contour for each segment from the values of $\bfR_i(t)$ using formulas from differential geometry \cite{orynyak2022,orynyak2025b}:
 \begin{gather}
    s(t)=\int_{0}^{t}{\sqrt{(\dot{x}(u))^2 + (\dot{y}(u))^2}\,du}\;, \\
    \kappa_d(t) = \frac{\dot{x}(t)\ddot{y}(t) - \dot{y}(t)\ddot{x}(t)}{((\dot{x}(t))^2 + (\dot{y}(t))^2)^{3/2}}\;.
\end{gather}

Obviously, having the dependencies of the parameters $s(t)$ and $\kappa_d(t)$, one can construct the dependency $\kappa(s)$.

\subsection{Adaptive Rigidities}

This section concerns the choice of rigidity $D_i$ in equation \eqref{eq:Dependences4} for the $\bfB_i$ points. It is intuitively clear that the density of measurement points can vary. Moreover, measurements often contain artifacts---they have locally distorted geometries, and thus correspond to a greater number of points per unit length of the smoothed contour. Certainly such areas should be smoothed more strongly than others. But when there are many ``noisy'' points, they are averaged, and their influence remains greater than where there are few noisy points. This is precisely why the concept of adaptive bandwidth is used in statistics \cite{eubank1999}.

Our adaptive rigidities are inspired by the idea of a beam on an elastic foundation (in an elastic medium) \cite{timoshenko1983} widely used in structural mechanics. Consider the following equation, which, unlike \eqref{eq:Dependences4}, applies to the entire segment, not just to discrete measurement points:
\begin{equation}
	\frac{dQ(t)}{dt} = -\frac{1}{h^4}(W(t) - \Pi(t))\;.
    \label{eq:Dependences4Midyfy}
\end{equation}

The function $\Pi(t)$ specifies the external excitation, i.e., it is a continuous external deviation. 
The general differential equation takes the form
\begin{equation}
	\frac{d^4W(t)}{dt^4} = -\frac{1}{h^4}(W(t) - \Pi(t)) = 0\;,
\end{equation}
where $h$ has the dimension of length and characterizes the joint interaction of the beam and the surrounding medium. The smoothing model (for an initially straight beam) \eqref{eq:Dependences4Midyfy} is interesting because it easily predicts the degree of smoothing for known deviation functions $\Pi(t)$. Let this function be given by
\begin{equation}
	\Pi\left(t\right)=\Delta \cos\frac{t}{\xi}\;,
\end{equation}
where $\Delta$ is the magnitude and $\xi$ is the semi-period of external excitation. Then the solution $W(t)$ to \eqref{eq:Dependences4Midyfy} is given by
\begin{equation}
	W(t) = \Delta \cos\frac{t}{\xi} \ep\left(\frac{\xi}{h}\right)\;,
\end{equation}
where the smoothing function (result) $\ep\left(\frac{\xi}{h}\right)$ is
\begin{equation}
	\ep\left(\frac{\xi}{h}\right) = \frac{\xi^4}{h^4 + \xi^4}\;.
\end{equation}

Obviously, if $\xi < h/2$, then $\ep(\frac{\xi}{h})\rightarrow 0$, and if $\xi > 2h$, then $\ep(\frac{\xi}{h})\rightarrow 1$. Therefore, knowing the length of the artifact $\xi$, one can select the smoothing parameter $h$ to minimize its influence. Conversely, knowing the length of local useful features, one can choose $h$ to preserve them. 

But our beam is on concentrated supports, not distributed ones. To preserve the properties of uniform smoothing as provided by the model of a beam in an elastic medium, we assume that the segment with an existing measured displacement is sufficiently small compared to $h$. Then in \eqref{eq:Dependences4Midyfy}, we can consider $W(t)$ and $\Pi(t)$ to be constants. Integrating, we obtain, instead of \eqref{eq:Dependences4}:
\begin{equation}
    Q_{i+1,0} = Q_{i,1} - \frac{L_i}{h^4} \left(W_{i,1} - \Pi_i\right)\;,
    \label{eq:Hardness}
\end{equation}
i.e.
\begin{equation}
    D_i = \frac{L_i}{h^4}\;,
\end{equation}
where $L_i$ is the length over which this support acts, i.e., the sum of half the distance to the right support and half the distance to the left support:
\begin{equation}
    L_i=\frac{s_i}{2}+\frac{s_{i+1}}{2}\;.
    \label{eq:SemiLen}
\end{equation}

In fact, applying the rigidity model \eqref{eq:Hardness} instead of \eqref{eq:Dependences4} means transitioning from a model of equal influence of each point to a model of equal influence of equal segments, regardless of how many points are on it. 

\section{Smoothing and Matching of Jigsaw Puzzles}

\subsection{Initial Contour and Point Renumbering}

\begin{figure}[tp]
	\centering
	\begin{subfigure}{0.45\textwidth}
		\centering
		\includegraphics[width=0.7\linewidth]{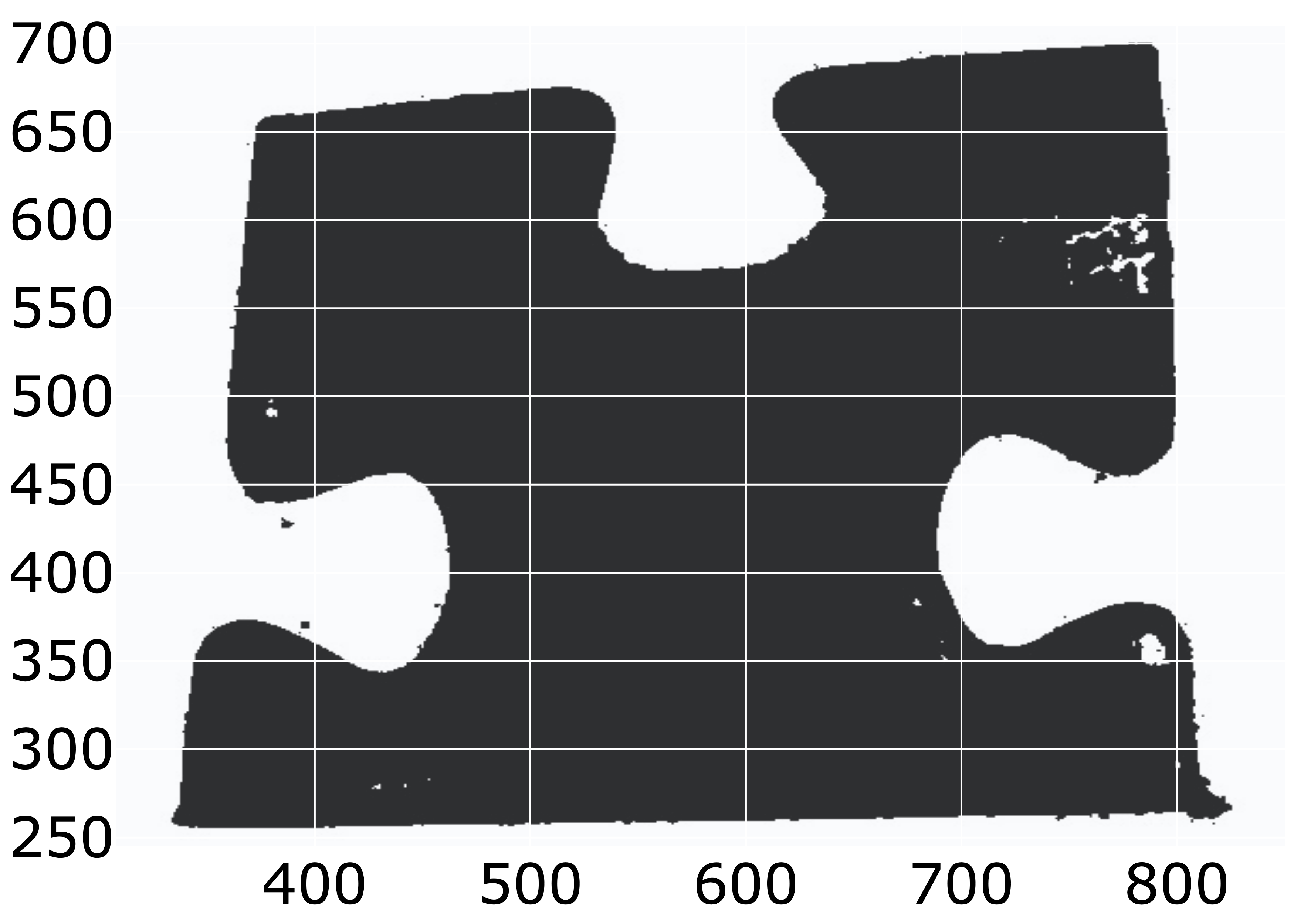}
		\caption{}
		\label{fig:Mask}
	\end{subfigure}\hfill
	\begin{subfigure}{0.45\textwidth}
		\centering
		\includegraphics[width=\linewidth]{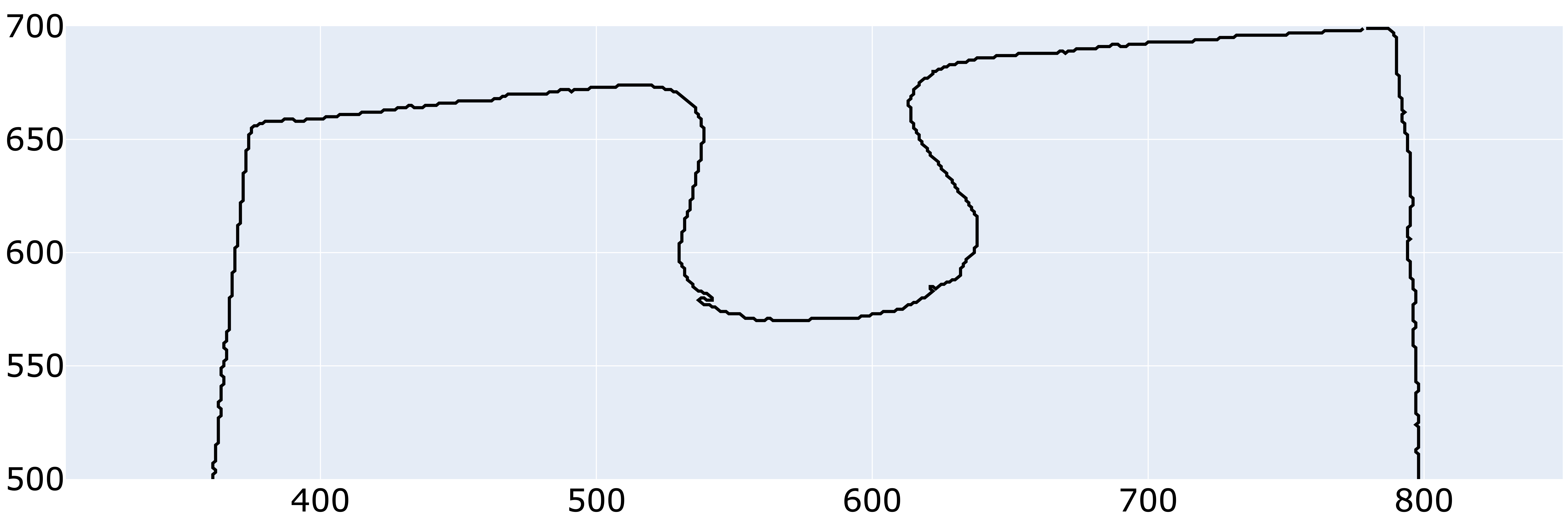}
		\caption{Renumbered $\bfB_i$ points according to the placement of their projections $\bfA_i$}
		\label{fig:Contour}
	\end{subfigure}\hfill
	\caption{(a) Puzzle piece mask and (b) $\bfB_i$ points in a puzzle piece}
\end{figure}

\begin{figure}[tp]
    \centering
    \includegraphics[width=0.6\textwidth]{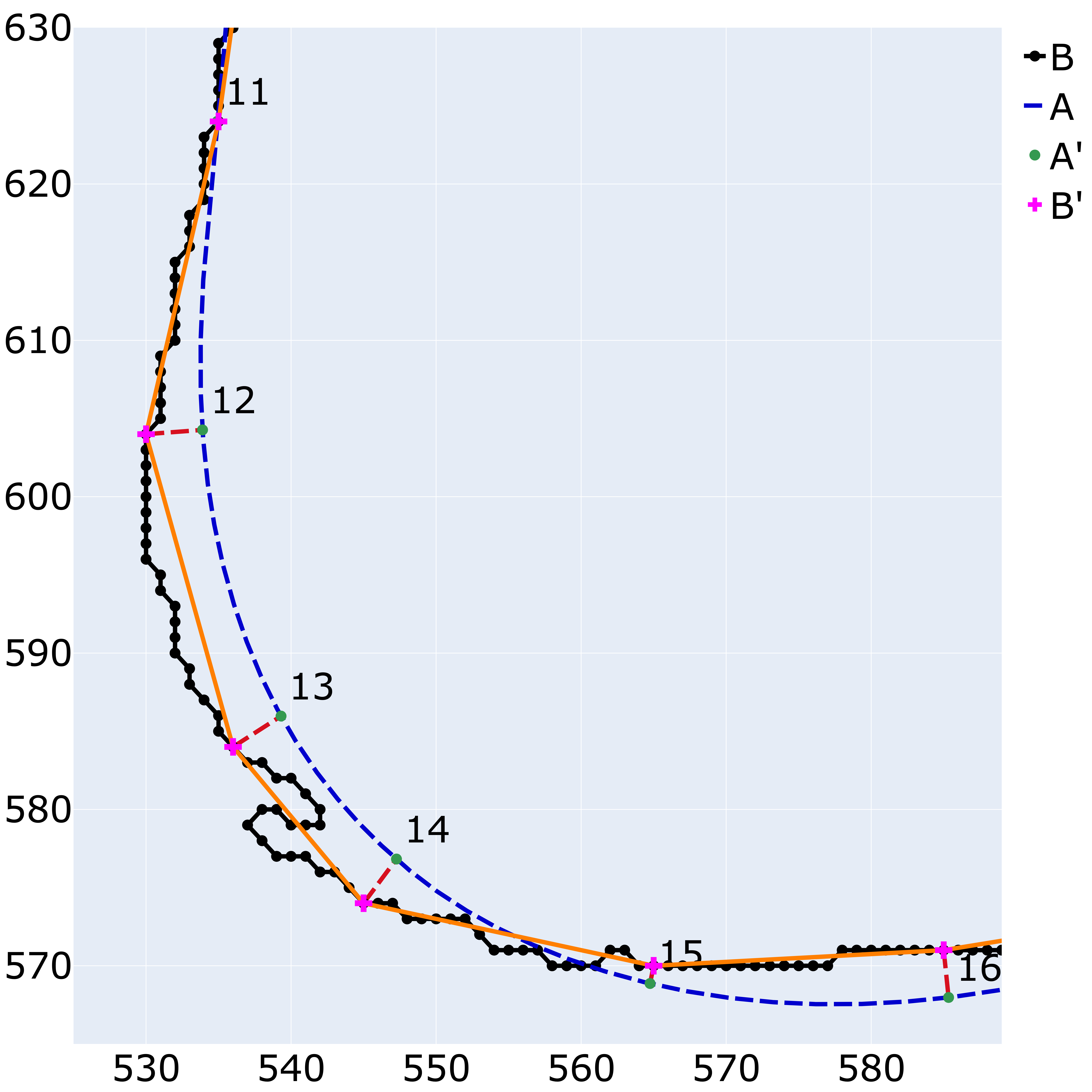}
    \caption{Contour after the first iteration. The black dots are the $\bfB_i$ points. The purple dots are the silhouette points (every 20th measured point). The orange line is the initial polygon, and the blue contour shows the calculated curve after 1 iteration. The red dashed lines show the correspondence between every silhouette  point and the points on the smoothed contour, i.e., their corresponding $\bfA_i$ points}
    \label{fig:PartFirstIter}
\end{figure}

\begin{figure}[tp]
    \centering
    \begin{subfigure}{0.45\textwidth}
        \centering
        \includegraphics[width=\linewidth]{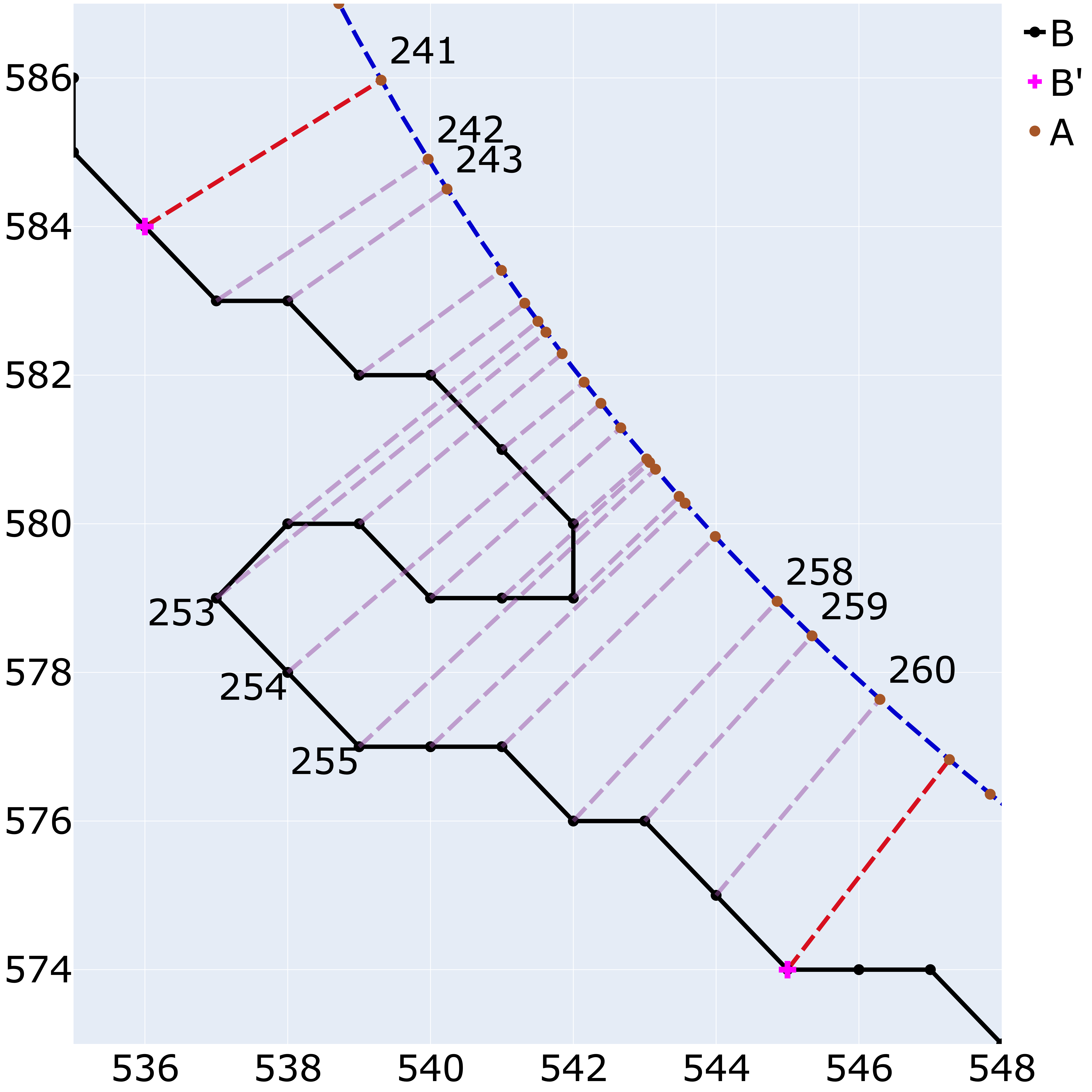}
        \caption{The $\bfB_i$ points according to the initial numbering and their projections (purple) onto the calculated contour}
        \label{fig:FitstIterOrder}
    \end{subfigure}\hfill
    \begin{subfigure}{0.45\textwidth}
        \centering
        \includegraphics[width=\linewidth]{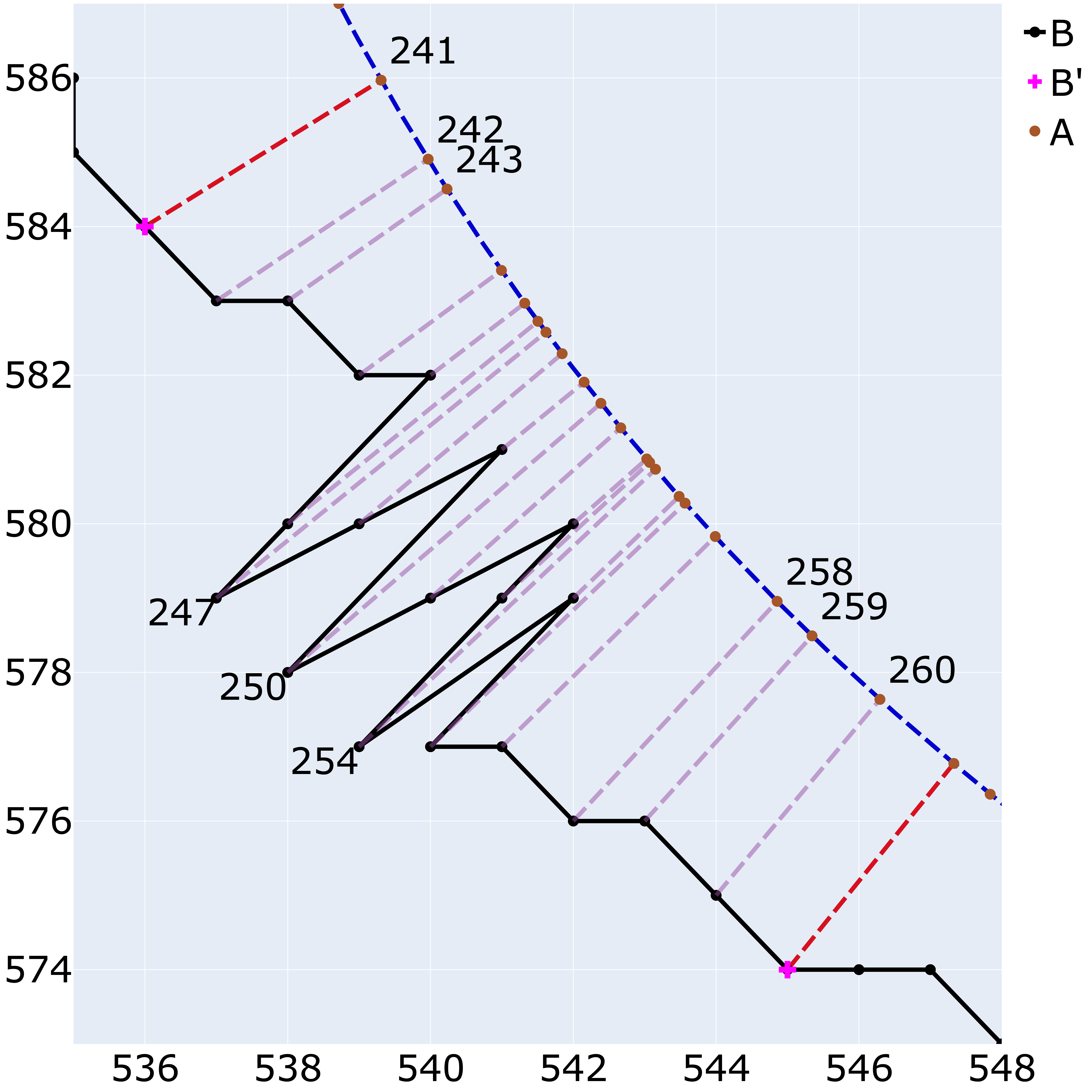}
        \caption{Renumbered $\bfB_i$ points according to the placement of their projections $\bfA_i$}
        \label{fig:FitstIterNewOrder}
    \end{subfigure}\hfill
    \caption{Projections of the $\bfB_i$ points onto the calculated contour and their renumbering}
\end{figure}

This section constitutes the fundamental methodological difference of this work from previous ones \cite{orynyak2025,orynyak2022,orynyak2025b} when it comes to curve smoothing. Let us discuss the main steps of the initialization procedure:
\begin{enumerate}
	\item We scan all the puzzle pieces to obtain a color model. We use the HSV (hue, saturation, value/brightness) model, which gives us more flexibility in color definition. To define the white color, a separation rule is introduced. For example, everything that falls within the range $(0, 0, 240)$--$(360, 50, 255)$ (all shades with a saturation value up to 50 and a brightness value from 240 to 255) is considered white, and everything else is black. As a result, we obtain a black and white image, where the puzzle piece is black and the background is white. Next, we invert the colors so that the background becomes black and the puzzle piece white,  obtaining a mask of the puzzle piece (Fig.~\ref{fig:Mask}). The \texttt{findContours} function from the OpenCV library is then applied to this image, which yields the initial points $\bfB_i$, shown for the upper part of the puzzle piece in Fig.~\ref{fig:Contour}.
	
	\item The corner points are identified. The identification of the 4 corner points is well-described in the literature, especially for standard quadrilateral elements, and takes into account that before an almost right angle (a sharp change in curvature), there are straight sections of sufficient length \cite{wolfson1988}. For each puzzle piece, the contour is divided into 4 separate sub-contours, each of whim has its own number. Subsequently, the splines are constructed separately for each sub-contour, with the following boundary conditions being applied at the ends of each sub-contour (left, $L$, and right, $R$):
	\begin{gather}
		W_L = M_L = 0\;, \\
		W_R = M_R = 0\;,
	\end{gather}
	which fix the position of the corner points and allow the slope angle at them to be arbitrary.
	
	\item In the first iteration, every $K$th point among the $\bfB_i$ points is selected, for example, every 20th, so that in the initial polygon, the angles $\psi_i$ from Fig.~\ref{fig:Model} are less than $90^\circ$.  We will call such points the silhouette  points. For this initial polygon, we assume that the $\bfA_i$ points coincide with the $\bfB_i$ points, which means the initial $\Pi_i$ are all zero. We choose an initial value of $h=K$ (i.e., it is assumed that the lengths of all artifacts are known to be less than $K$). We compute the first iteration, after which we obtain an initial approximate contour that we will refine further by taking into account all other points and refining (decreasing) the value of $h$. For a local region, we will obtain the calculated (blue) contour, as shown in Fig.~\ref{fig:PartFirstIter}.
	
	\item Our task now is to find the projections of all $\bfB_i$ points without exception onto the contour and obtain the corresponding $\bfA_i$ points. Moreover, even those $\bfB_i$ points that participated in the construction of the approximate contour (silhouette points), for which we already have corresponding calculated values $\bfA_i$, also take part in the refining procedure, which consists of redefining and renumbering of the $\bfA_i$ points, and, consequently, the renumbering of the $\bfB_i$ points. For this, 200 candidate points, equidistantly placed between the silhouette  points, are considered, and the nearest distances between the real $\bfB_i$ points and these candidate points are calculated (within the 3 nearest ``silhouette  segments''). Thus, new $\bfA_i$ points are found, which are numbered according to their placement along the length of the contour, and then the $\bfB_i$ points are renumbered accordingly. Figure~\ref{fig:FitstIterOrder} shows the $\bfB_i$ points connected according to the initial numbering, and Fig.~\ref{fig:FitstIterNewOrder} shows them connected according to the obtained renumbering.
\end{enumerate}

Next, the spline refinement proceeds iteration by iteration as follows. Having, at iteration $k$, the sequence of $\bfA_i^k$ points, which correspond to and are locally closest to $\bfB_i$ points, and thus the distances $\Pi_i$, we choose a new value for the smoothing parameter, for example, according to the rule $h^{k+1} = h^k/1.2$, which provides less smoothing compared to the previous one. We construct a new polygon from the $\bfA_i^k$ points, build a spline, and find a new contour. We find the $\bfA^{k+1}_i$ points as the points on the contour, closest to the $\bfB_i$ points (using the same approach as above). If necessary, we again renumber the $\bfA^{k + 1}_i$ to ensure that they are numbered according to their (spline arc length) coordinate $s$. The $\bfB_i$ points are renumbered accordingly.

\subsection{Matching Criteria}

To determine the quality of a curve, the following well-known formula is used:
\begin{equation}
	E = \int_{0}^{l}{\kappa^2(s) \, ds}\;,
    \label{eq:Criterion}
\end{equation}
where $\kappa(s)$ is the curvature. The smaller the obtained energy value $E$, the better the quality of the curve. Equation \eqref{eq:Criterion} is used for the initial analysis of each side of each contour. In particular, a certain criterion value $E_{st}$ is fixed, by which straight (puzzle border) and curved sides are separated.

To compare two sides of a puzzle piece, we apply \eqref{eq:Criterion} to the corresponding curvatures (one of which is calculated in the opposite direction) as follows:
\begin{equation}
	E_{a,b} = \int_{0}^{l} \left(\kappa_1(s)-\kappa_2(s)\right)^2\,ds\;.
    \label{eq:SidesCriterion}
\end{equation}
where we have introduced the concept of comparison energy (criterion) $E_{a,b}$ as the energy of the mutual difference in curvatures of two different sides $a$ and $b$. Thus, we find the quality value of the difference in curvatures. The sides of different puzzle elements are considered (candidates for) matching if their mutual energy yields the smallest value. 

\section{Illustrative Examples}

\subsection{Jigsaw Puzzle Description}

\begin{figure}[tp]
    \centering
    \includegraphics[width=0.5\textwidth]{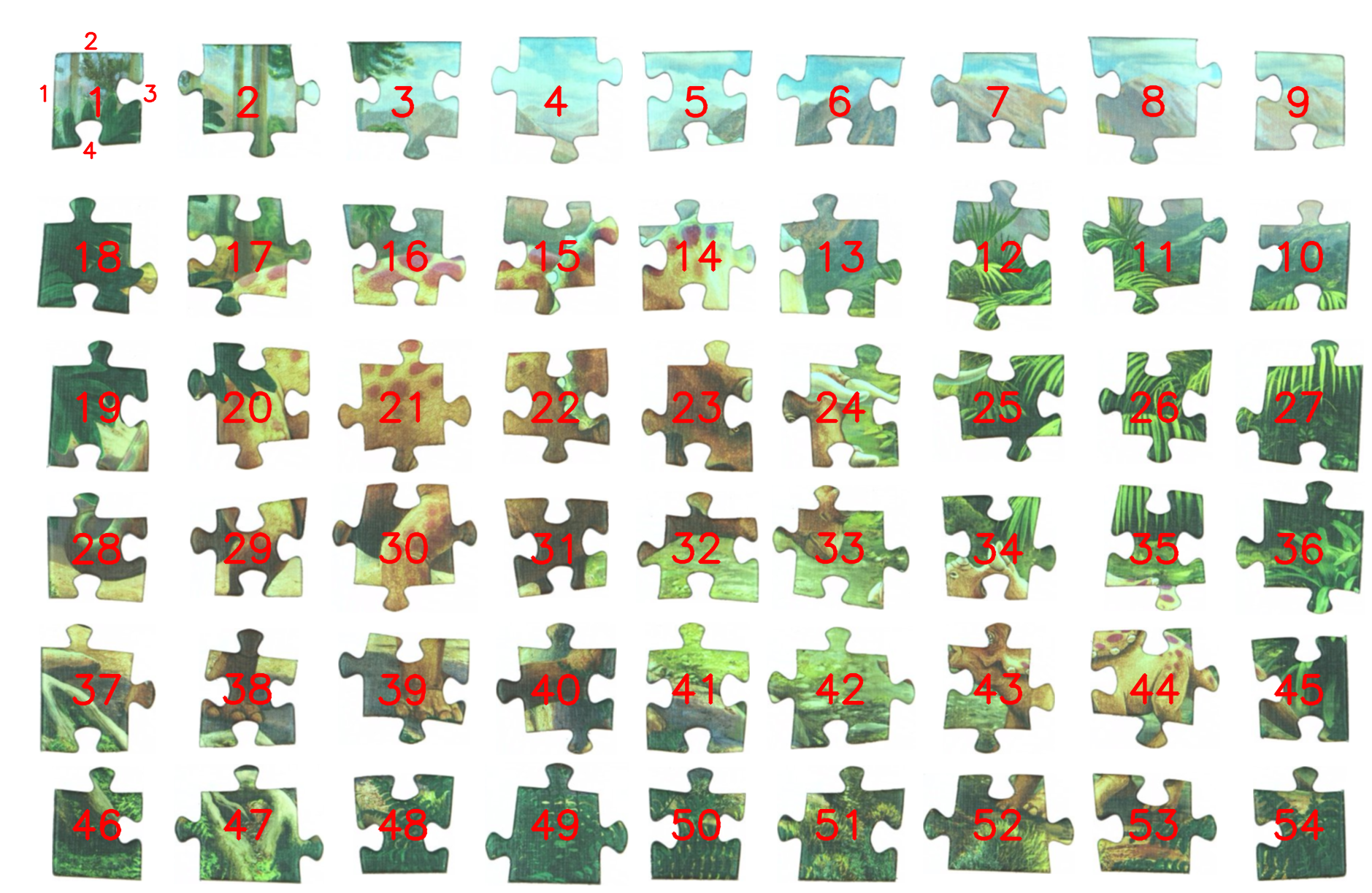}
    \caption{The puzzle under consideration}
    \label{fig:PuzzlePieces}
\end{figure}

The 54-piece puzzle under consideration is shown in Fig.~\ref{fig:PuzzlePieces}. All elements are numbered. Each element has 4 sides, which are also numbered clockwise. For puzzle piece No. 1, the order of side numbering is shown. Hereafter, the sides of the puzzle pieces will be denoted by the puzzle number and the side number, separated by a hyphen. Thus, 1-1 is the left side of the first puzzle piece.

Thus, we have $54\cdot 4 = 216$ sides. Each side must be matched with another. Of course, the presence of corner and straight sides, as well as convex and concave sides, significantly reduces the number of necessary comparisons, as discussed in detail in \cite{goldberg2002,wolfson1990,goldberg2004}. Our task is to match the curvatures of the corresponding sides.

\subsection{Matching Sides}

Let us describe the methodology for matching two contours. We determine the curvatures and the total lengths of each side. They may differ slightly due to measurement inaccuracies (different photographing distances, differences in the accuracy of corner points determination). To weed out the pairs that obviously do not match with each other, we apply a separation rule---if the length ratio is less than, say, 97\%, the sides are considered non-matching. For the remaining pairs, we rescale the shorter side to the longer one (we introduce scaling of the abscissa $s$) so that their lengths are the same. This rescaling rule is applied separately to each pair of puzzle pieces we compare. 

A shift rule is also applied, which reflects the fact that corner points may be determined incorrectly, by 2-4 pixels in different directions. Therefore, when comparing two contours, the first one is considered in several shifted positions relative to the second, i.e., it is sequentially shifted by $4$, $2$, $0$, $-2$, and $-4$ pixels (correspondingly, the $\bfB_i$ points are excluded). The energies are calculated, and the smallest among all energies is chosen.

As an example of side matching, let us consider side 47-2 of element No.~47 (Fig.~\ref{fig:PuzzlePieces}) as one of the most problematic in assembling the entire puzzle. The fact is that the correct matching side, namely 38-4, and one of the ``incorrect'' sides, namely 22-2, gave very close results in terms of matching energy. Therefore, it is interesting to examine them in more detail. Let us investigate how the smoothing parameter $h$ affects the obtained values of the comparison energy \eqref{eq:SidesCriterion}. We will fix the value of $h$ and perform calculations for both pairs. First, let's compare the lengths of the sides (Table~\ref{tab:Scales}). As we can see, small inaccuracies in scaling and the initial position of the puzzle pieces during photography slightly affect the ratio of side lengths and change insignificantly with parameter $h$. This table gives an idea of the inaccuracy of length measurement and justifies the criterion for rejecting pairs by length, which we have set here with a margin at 97\%.

\begin{table}[tp]
\centering
\caption{Comparison of length ratios of 47-2 with 38-4 and 47-2 with 22-2, for different $h$}
\begin{tabular}{c|c|c}
\hline
\multirow{2}{*}{\ } & \multicolumn{2}{c}{Pairs of sides} \\
\cline{2-3}
 & 47-2 and 38-4 & 47-2 and 22-2 \\ \hline
5 & 98.75 & 97.12 \\ \hline
8 & 98.77 & 97.1 \\ \hline
10 & 98.77 & 97.1 \\ \hline
12 & 98.76 & 97.1 \\ \hline
15 & 98.71 & 97.05 \\ \hline
\end{tabular}
\label{tab:Scales}
\end{table}

Let us now move on to the matching criterion itself. The energy of side 47-2 calculated using \eqref{eq:Criterion} is given in Table~\ref{tab:Energy}. As we can see, the larger the value of $h$, the lower the deformation energy of the contour, which is logical. This energy changes from $0.30003$ for $h=5$ to $0.22096$ for $h=15$. These values give an idea of the number against which the matching energies of the contours will be compared. As we can see, for $h=5$, the matching energy of contours 47-2 and 38-4, and 47-2 and 22-2 is 7.23\% and 12.86\% of the energy of 47-2 alone, respectively. It is interesting to note two smoothing trends at different values of $h$. First, the ratio of the matching energy to the initial energy decreases (i.e., the relative correspondence improves). Second, the ratio between the energies of the ``correct'' pair and the ``incorrect'' pair reaches its maximum at $h=12$. That is, for $h=12$, the matching energies of contours 47-2 and 38-4, and of contours 47-2 and 22-2, are 0.41\% and 2.32\%, respectively. We emphasize again that these two pairs of contours (``correct'' and ``incorrect'') were the most problematic in our puzzle.

\begin{table}[tp]
\centering
\caption{Comparison of sides: 47-2 only, 47-2 with 38-4, and 47-2 with 22-2, for different $h$, expressed in energy values}
\begin{tabular}{c|c|c|c}
\hline
& Only 47-2 & 47-2 and 38-4 & 47-2 and 22-2 \\ \hline
5 & 0.300030 & 0.021703 & 0.038588 \\ \hline
8 & 0.271185 & 0.004577 & 0.013564 \\ \hline
10 & 0.257161 & 0.002128 & 0.008786 \\ \hline
12 & 0.244003 & 0.001014 & 0.005684 \\ \hline
15 & 0.220964 & 0.000368 & 0.003109 \\ \hline
\end{tabular}
\label{tab:Energy}
\end{table}

 \begin{figure}[tp]
    \centering
    \begin{subfigure}{\textwidth}
        \centering
        \includegraphics[width=\linewidth]{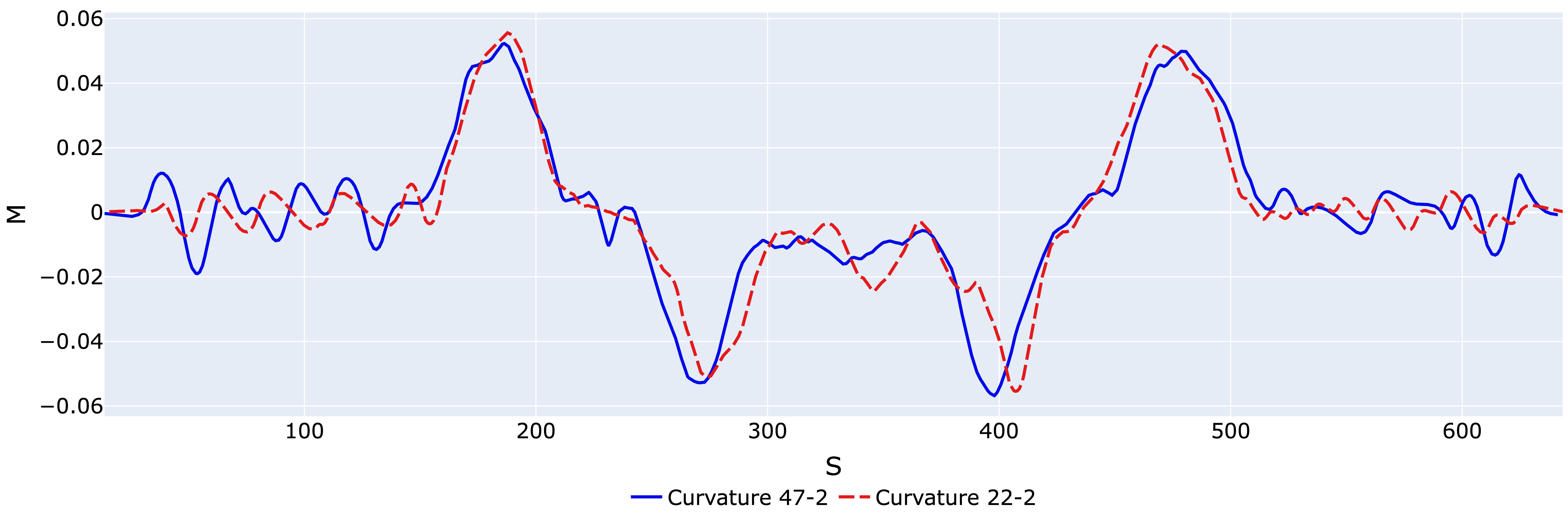}
        \caption{47-2 and 38-4 }
        \label{fig:H5_47_38}
    \end{subfigure}\hfill
    \begin{subfigure}{\textwidth}
        \centering
        \includegraphics[width=\linewidth]{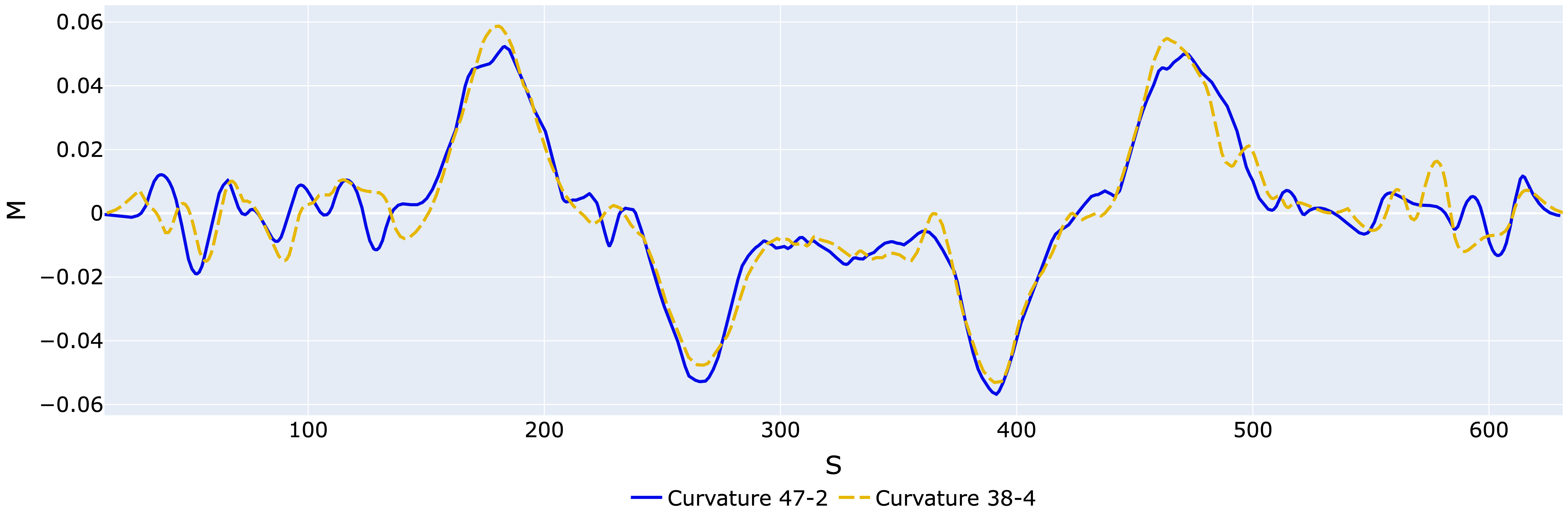}
        \caption{47-2 and 22-2}
        \label{fig:H5_47_22}
    \end{subfigure}
    \caption{Comparison for $h=5$}
\end{figure}

 Let us illustrate these results graphically. Figure~\ref{fig:H5_47_38} shows the curvature comparisons for 47-2 and 38-4 at $h=5$. It is evident that the curvature plots are quite rough, have many local peaks, and appear under-smoothed. This is precisely why the energy value is quite large. Similarly, curvature plots are provided for the pair of contours 47-2 and 22-2 (Fig.~\ref{fig:H5_47_22}). Visually, it is difficult to say which pair is better, although formally the energy value for 47-2 and 38-4 is better, albeit only slightly (Table~\ref{tab:Energy}).

%

\begin{figure}[tp]
    \centering
    \begin{subfigure}{\textwidth}
        \centering
        \includegraphics[width=\linewidth]{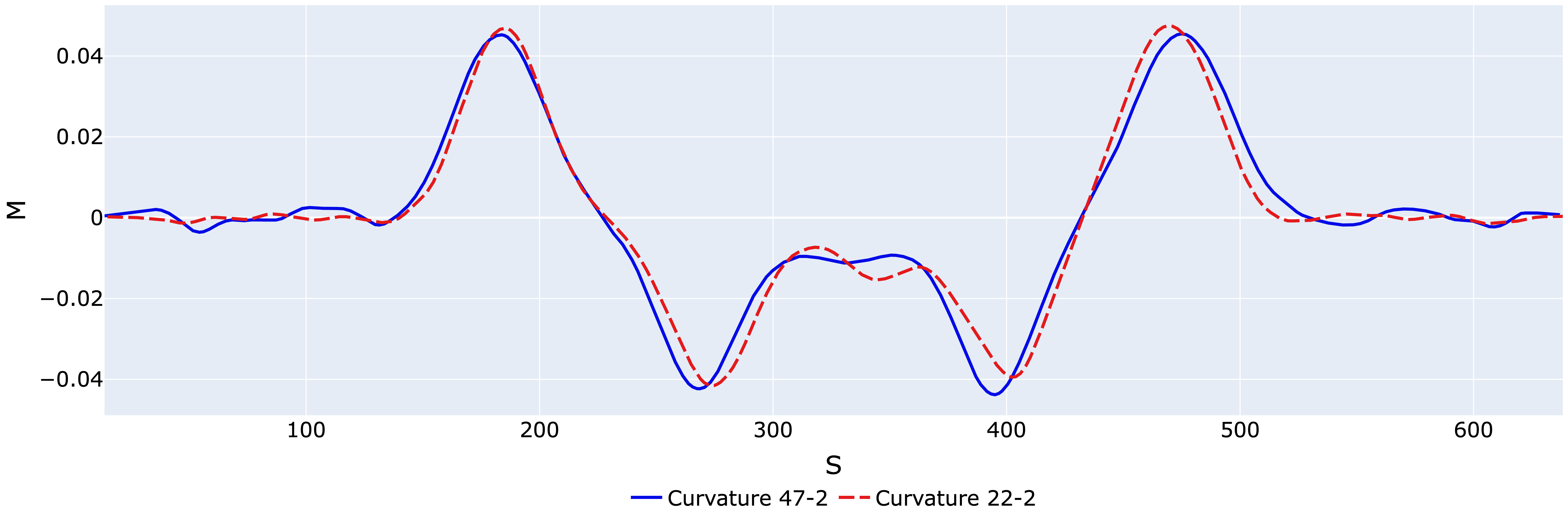}
        \caption{47-2 and 38-4}
        \label{fig:H10_47_38}
    \end{subfigure}\hfill
    \begin{subfigure}{\textwidth}
        \centering
        \includegraphics[width=\linewidth]{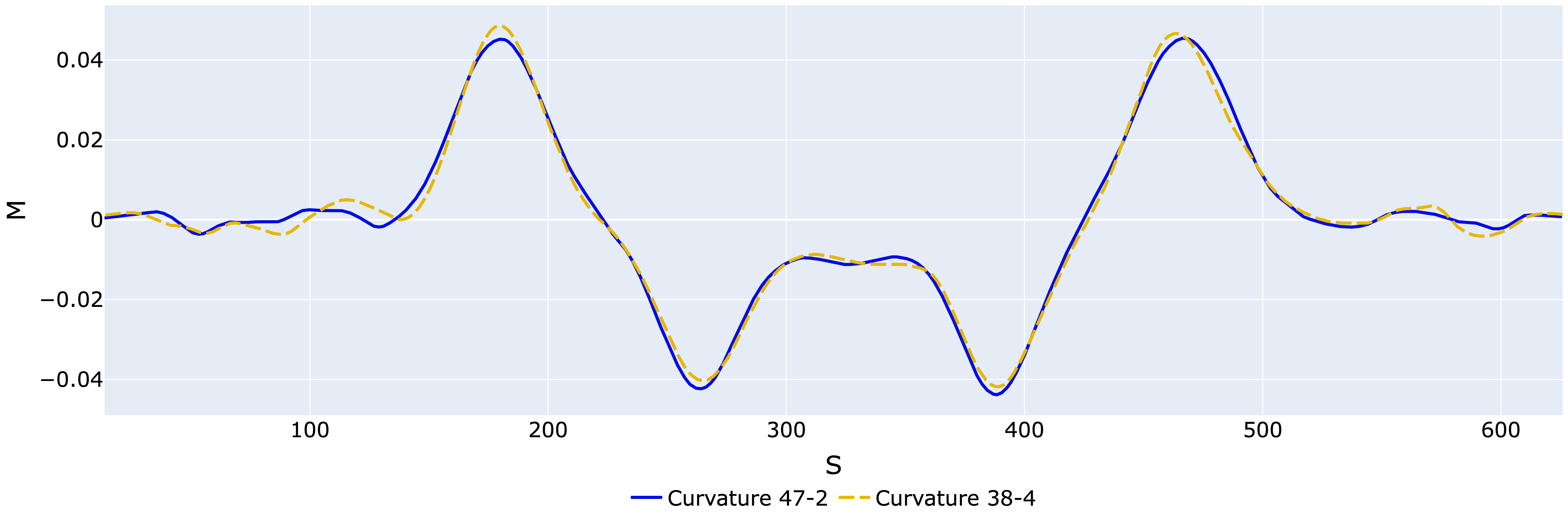}
        \caption{47-2 and 22-2}
        \label{fig:H10_47_22}
    \end{subfigure}\hfill
    \caption{Comparison for $h=10$}
    \label{fig:H10}
\end{figure}

Using $h=10$ further noticeably reduces the comparison energy, as shown in Fig.~\ref{fig:H10}. The ``correct'' pair has better energy indicators (almost 4 times better) and visually looks better than the ``incorrect'' pair. Even the behavior of the graphs at the edges is more logical, as the lines are almost straight. This indicates a sufficient level of smoothing.


 \begin{figure}[tp]
    \centering
    \begin{subfigure}{\textwidth}
        \centering
        \includegraphics[width=\linewidth]{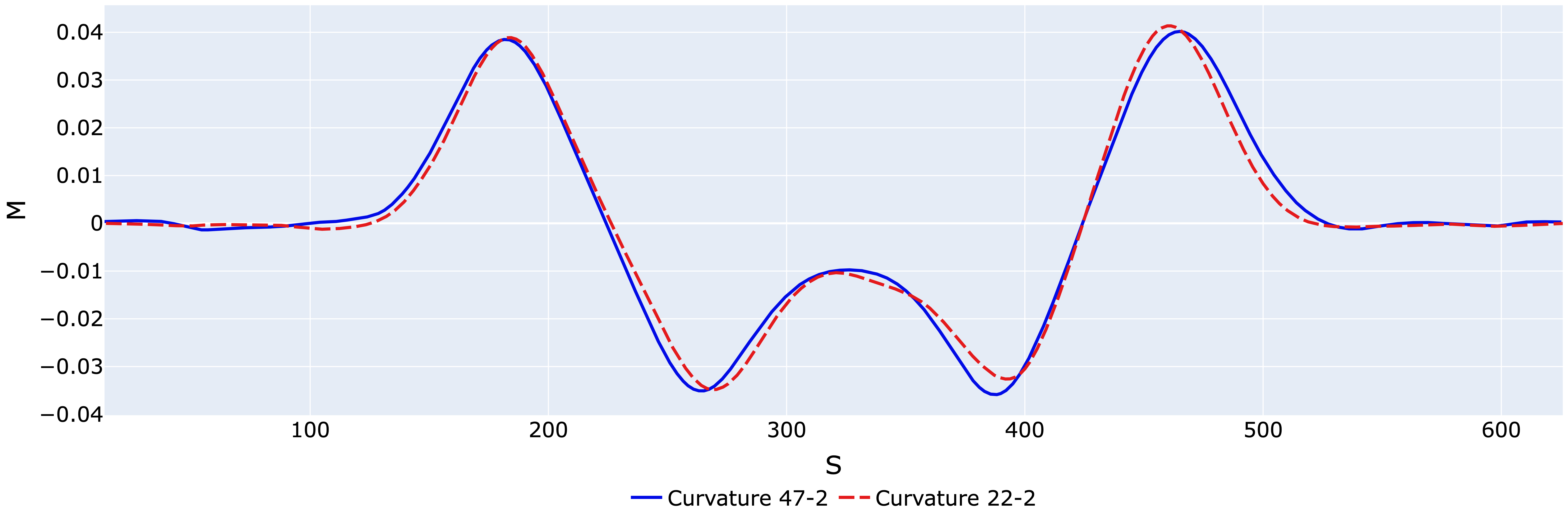}
        \caption{47-2 and 38-4}
        \label{fig:H15_47_38}
    \end{subfigure}\hfill
    \begin{subfigure}{\textwidth}
        \centering
        \includegraphics[width=\linewidth]{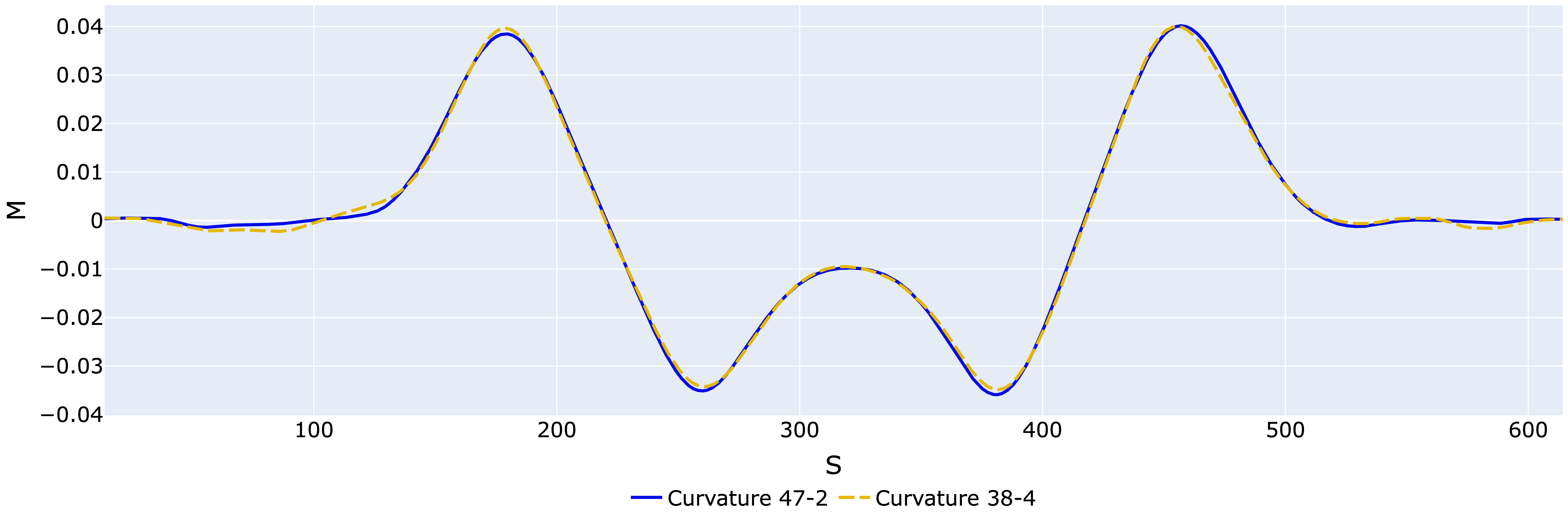}
        \caption{47-2 and 22-2}
        \label{fig:H15_47_22}
    \end{subfigure}\hfill
    \caption{Comparison for $h=15$}
    \label{fig:H15}
\end{figure}

Further smoothing at $h=15$ again improves the results (Fig.~\ref{fig:H15}). But unlike the previous results, it can be seen that the maximum curvatures (in absolute value) have decreased from $0.042$ to $0.035$, which is evidence of some over-smoothing. The energy has decreased again, especially significantly for the correct pair. Now the comparison energy is very small for both cases and is only $0.17$\% of the initial energy of the contour. The ``correct'' energy value is 8.5 times smaller than the ``incorrect'' one, but the absolute values are very small.

\begin{figure}[tp]
    \centering
    \includegraphics[width=\textwidth]{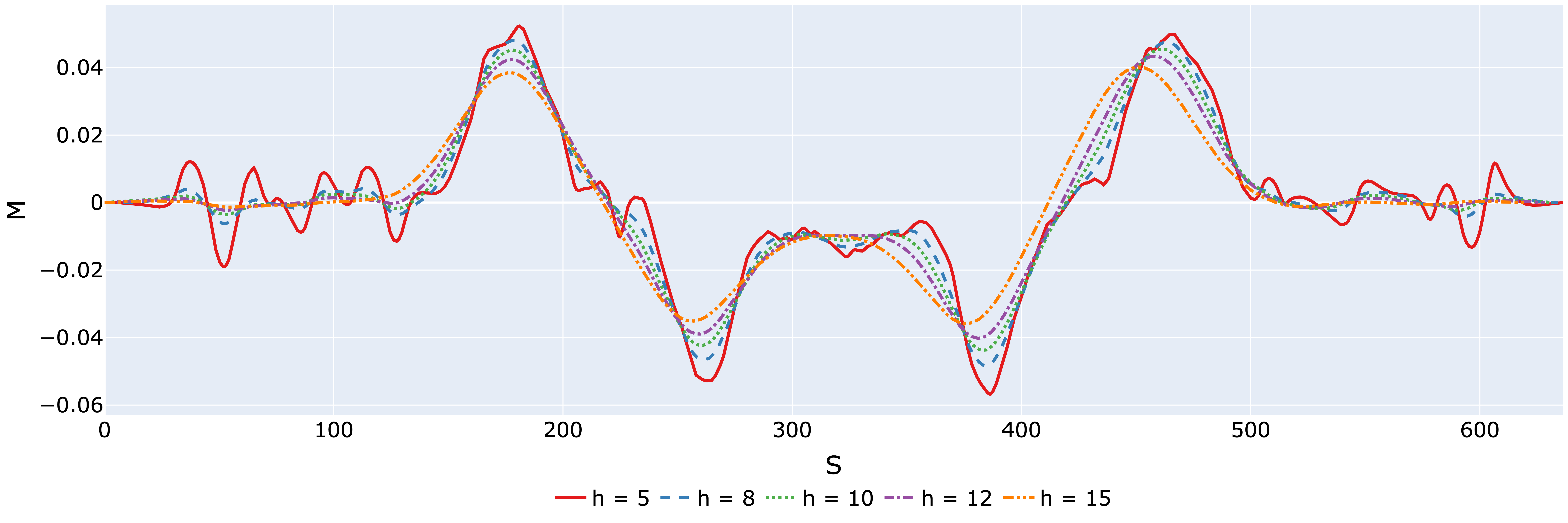}
    \caption{The puzzle under consideration. Graph comparing curvatures at different values of $h = 5, 8, 10 ,12, 15$}
    \label{fig:AllH}
\end{figure}

To illustrate the effect of smoothing on curvatures, we show a graph of the curvature of contour 47-2 at different $h$ (Fig.~\ref{fig:AllH}). Here, all the smoothing phenomena described above are more clearly observed. What is important for us is that the other contour (38-4) is also smoothed similarly, meaning that despite the first contour changing both its extrema and their locations, the second contour behaves in a similar manner. This means that the methodology for choosing the concept of $h$ is correct.

\subsection{Effect of Positional Shifts}

Let us demonstrate the effect of shifting on calculating the energy for the pair 47-2 and 38-4. Since the corners and vertices may not be determined with perfect accuracy, we apply a shift algorithm everywhere. For the first side, we take 4 points in both directions at a distance of approximately 4 pixels. We check 5 shift combinations (4 to the left, 2 to the left, no shift, 2 to the right, and 4 to the right). The data are presented in Table~\ref{tab:Shift_47_38}, where the best values are highlighted in bold.

 \begin{table}
\centering
\caption{Effect of shift on the energy value for the pair of sides 47-2 and 38-4}
\begin{tabular}{c|c|c|c|c|c}
\hline
Shift & $-4$ & $-2$ & $0$ & $2$ & $4$ \\ \hline
5 & 0.044029 & 0.029984 & \textbf{0.021703} & 0.022939 & 0.031702 \\ \hline
8 & 0.016018 & 0.008011 & \textbf{0.004577} & 0.006153 & 0.012428 \\ \hline
10 & 0.01041 & 0.004515 & \textbf{0.002128} & 0.003456 & 0.008356 \\ \hline
12 & 0.007342 & 0.002743 & \textbf{0.001014} & 0.002244 & 0.006382 \\ \hline
15 & 0.004963 & 0.001552 & \textbf{0.000368} & 0.001439 & 0.004752 \\ \hline
\end{tabular}
\label{tab:Shift_47_38}
\end{table}

Similar results are shown for the pair 47-2 and 22-2 in Table~\ref{tab:Shift_47_22}. As we can see, within the specified limits, the shift has an insignificant effect on the energy. This means that the accuracy of determining the corner point within $\pm 3$ pixels has almost no effect on the results.

\begin{table}
\centering
\caption{Effect of shift on the energy value for the pair of sides 47-2 and 22-2}
\begin{tabular}{c|c|c|c|c|c}
\hline
Shift & $-4$ & $-2$ & $0$ & $2$ & $4$ \\ \hline
5 & 0.061865 & 0.048843 & 0.041013 & \textbf{0.038588} & 0.041525 \\ \hline
8 & 0.03014 & 0.020633 & 0.015029 & \textbf{0.013564} & 0.016567 \\ \hline
10 & 0.019771 & 0.012912 & 0.009309 & \textbf{0.008786} & 0.012476 \\ \hline
12 & 0.01365 & 0.008365 & \textbf{0.005684} & 0.006281 & 0.009607 \\ \hline
15 & 0.008928 & 0.004952 & \textbf{0.003109} & 0.003706 & 0.006481 \\ \hline
\end{tabular}
\label{tab:Shift_47_22}
\end{table}

\subsection{Jigsaw Puzzle Reconstruction}

The first step in assembling the jigsaw puzzle is to calculate energies of each side and sort the sides into straight and curved ones. The curved ones can optionally be also sorted into convex and concave. It is an easy task as the maximal energy of a straight side was dozens of times higher than the minimal energy of any curved one. At $h = 10$, the minimal energy for a curved side is $0.2368$, whereas the maximal energy of a straight side is only $0.00118$. Thus, we designated any side with energy lower than $0.1$ as a straight one. This threshold remains the same for all other values of $h$.

We start the process with a corner piece. In our case, there are four corner pieces, and any of them can be chosen. We treat a piece as a corner one if it has two straight sides with the smallest sum of energies \eqref{eq:Criterion}. According to this criterion, we selected piece No. 54, for which at $h = 10$ we have $E_{54-3}=0.000169$ and $E_{54-4}=0.000989$.

The assembly process commences. The side chosen as the first to be matched was the curved side that follows the two straight sides in clockwise order. Thus, we begin by matching side 54-1. Since piece No. 54 is a corner piece, it must connect to another border piece. Accordingly, we select as potential candidates all pieces that have at least one straight side. This includes other corner pieces, as the overall dimensions of the puzzle (number of rows and columns) are not yet known. Among these candidates, we only consider sides that precede a straight side in clockwise order and are convex (since side 54-1 is concave), such as sides 53-3, 4-1, or 36-2. For side 54-1, the best match was found to be side 53-3, with a matching energy of $E_{54-1, 53-3}=0.002034$ and a length ratio of 99.62\%. The next best candidate was side 10-2, which had a length ratio of 98.02\% but a joint energy of $E_{54-1, 10-2}=0.149664$. This energy value is nearly 75 times greater than that of the correct match, demonstrating the powerful discriminative capability of the energy metric.

Following this greedy selection principle, we fill the entire first row until another corner piece is found. 
 The next corner piece identified was piece No. 46. At this point, we have established that the puzzle has 9 columns. Knowing the total number of pieces, we can now deduce the number of rows.


After completing the first row, we change direction and search for the piece that fits above the most recently placed corner piece. Since this is also a border piece, we search using the same principle (but excluding the already-placed corner pieces from the candidate set). In our case, this piece is No. 37. 


Next, we continue to assemble the second row, proceeding in the opposite direction (left-to-right). However, since we are now placing an interior piece, not a border piece, the search for the next fragment is constrained by two curved sides. The piece being sought must match the already-placed sides 47-2 and 37-3 (in terms of both length and energy) in that clockwise order, and it must have concave shapes for both corresponding sides. This two-sided constraint immediately prunes the vast majority of pieces and their orientations from the search space, leaving only 6 valid combinations, such as 22-(3,2) and 38-(1,4). For the correct match, piece No. 38, the comparison of curvatures yields joint energies of $E_{47-2,38-4}=0.002128$ and $E_{37-3,38-1}=0.0014$. The sum of these energies is $0.003528$. The next-best combination, piece No. 22 in orientation (3,2), yields energy values of $E_{47-2,22-2}=0.008786$ and $E_{37-3,22-3}=0.021298$, for a total sum of $0.030084$. This sum is almost 8.5 times greater than the sum for the correct piece, No. 38. 
This outcome demonstrates the profound effectiveness of the energy criterion when applied to properly smoothed contours and their calculated curvatures.

%
%

The remainder of the assembly process is trivial. At no stage of the process were we faced with an ambiguous choice for the next piece or orientation to select for placement. For more complex puzzles, it might happen that energy criterion alone cannot be used to break ties between candidate puzzle pieces, and more complicated assembly strategies should be employed.

\section{Conclusion}

This work addresses the problem of curve matching, a fundamental challenge with significant applications in diverse fields such as computer graphics, forensic science, and pattern recognition. The task of solving jigsaw puzzles serves as an illustrative case study for the application of CBS to such problems. The main achievement of this work is the development of a method for processing dense point sets where the measurement error can exceed the distance between adjacent points. The primary contribution is to demonstrate that proper contour smoothing and accurate curvature calculation are of paramount importance. When these prerequisites are met, simple assembly algorithms can be effectively employed. In this paper, we have established the following:

\begin{enumerate}
	\item We justify the selection of an individual rigidity for each support point. This ensures that segments are smoothed uniformly according to their length, rather than the number of sample points. This rigidity is characterized by the effective length of the smoothing influence (analogous to bandwidth) and the distance between the projections of the sample points onto a trial contour.
	
	\item We introduce the concept of an initial (or starting) contour, which is constructed from a limited number of real points. This contour serves for the initial re-numbering of the data points.
	
	\item We propose a new method for constructing the $\bfA_i$ contour points, against which the $\bfB_i$ points are numbered. In contrast to \cite{orynyak2022}, we refine the positions of the $\bfA_i$ points after calculating the approximate contour in the preceding iteration. They are found as the nearest points to the $\bfB_i$ points within a local neighborhood of the previously found $\bfA_i$ points.
	
	\item A quantitative technique for comparing two contours is provided, based on calculating the integral of the squared difference of their curvatures.
	
	\item We investigated the technical nuances of the matching process, including shift, scaling, and the smoothing parameter $h$. While these factors did not play a significant role in the present case, they may be essential for more complex puzzles, for example, when the endpoints of the matching sides are unknown.
\end{enumerate}

\section*{Funding Sources}

This research did not receive any specific grant from funding agencies in the public, commercial, or not-for-profit sectors.

\bibliography{references}

\end{document}